\documentclass[showpacs,aps,graphicx,twocolumn]{revtex4}
\usepackage{txfonts}

\usepackage{graphicx}

\begin{document}

\title{Universal quantum gates on microwave photons assisted by circuit quantum
electrodynamics\footnote{Published in Phys. Rev. A \textbf{90},
012328 (2014)} }

\author{Ming Hua,  Ming-Jie Tao, and  Fu-Guo Deng\footnote{Corresponding author: fgdeng@bnu.edu.cn}}

\address{Department of  Physics, Applied Optics Beijing Area Major Laboratory, Beijing
Normal University, Beijing 100875, China}

\date{\today }

\begin{abstract}
Based on a microwave-photon quantum processor with two
superconducting resonators coupled to one transmon qutrit, we
construct the controlled-phase (c-phase) gate on
microwave-photon-resonator qudits, by combination of the
photon-number-dependent frequency-shift effect on the transmon qutrit
by the first resonator and the resonant operation between the qutrit
and the second resonator. This distinct feature provides us  a
useful way to achieve the c-phase gate on the two  resonator
qudits with a higher fidelity and a shorter operation time, compared
with the previous proposals. The fidelity of our c-phase gate can
reach $99.51\%$ within 93 ns. Moreover, our device can be extended
easily to construct the three-qudit gates on  three resonator qudits, far different from the existing proposals.
Our controlled-controlled-phase gate on three resonator
qudits is accomplished with the assistance of a transmon qutrit and its fidelity  can reach $92.92\%$ within 124.64 ns.
\end{abstract}

\pacs{ 03.67.Lx,  03.67.Bg,  85.25.Dq,  42.50.Pq} \maketitle


\section{Introduction}\label{sec1}

Quantum computation has attracted much attention in recent years
\cite{book}. Many important schemes have been proposed for quantum
computation by using different quantum systems,  such as photonic
systems \cite{photon1,photon2,photon3,photon4,photon5}, nuclear
magnetic resonance \cite{longjcp,longprl},  quantum dots
\cite{QD1,QD2,QD3,QD4}, and diamond nitrogen-vacancy (NV) centers
\cite{NV1,NV2,NV3}. Universal quantum gates  are the key elements in
constructing a universal quantum computer. Moreover, they can be
used to produce the entanglement of multipartite quantum systems.
The controlled-phase (c-phase) gate is one of the important universal
two-qubit gates. It has the same role as
the controlled-not gate in quantum computation. The controlled-controlled-phase (cc-phase) gate is an important
three-qubit gate which can play the same role as the three-qubit
Toffoli gate which can be used to construct a universal quantum
computation with single-qubit Hadamard operations \cite{book}.

Circuit quantum electrodynamics (QED), which combines
superconducting circuits and cavity QED, provides a good platform
for  quantum computation
\cite{Wallraff,SQ,liuyx,longglpra,Chiorescu}. A
superconducting Josephson junction can act as a perfect qubit and it
has some good features, such as the large scale integration
\cite{Lucero}, a relatively long coherence time of about $0.1$ ms
\cite{Rigetti}, the  versatility in its energy-level structure with
$\Xi $, $\Lambda $, $V$, and even $\Delta $ types \cite{You} which
cannot be found in  atom systems, and the superconducting qubit with
tunable coupling  strength \cite{Laloy,Harris,Srinivasan}. All these characteristics have attracted much attention focused on quantum
information processing on superconducting qubits in circuit QED.
Some interesting proposals for quantum information processing on
qubits have been  presented, such as the reset of a superconducting qubit
\cite{reset1,MDReed,reset6}, universal quantum gates and
entanglement generation \cite{DiCarlo,gate6,DiCarlo1,3q,3q1,error}, and
single-shot individual qubit measurement and  the  joint  qubit
readout \cite{measure1,measure2,JMChow}.

A superconducting coplanar resonator whose quality factor $Q$ can be
increased to be $10^{6}$ \cite{q1,q2,q3,Devoret}, can act as a qudit
because it contains some microwave photons whose lifetimes are much
longer than that of a superconducting qubit
\cite{Devoret,longer,longer1}. The coupling strength between a
resonator and a transmission line is tunable \cite{Yin}. Moreover,
the strong and even ultrastrong coupling
\cite{SQ,Forn,strong,strong1,strong2,strong3,strong4}  in circuit
QED affords a strong nonlinear interaction between a superconducting
qubit and a microwave-photon qudit. These good features make
resonators  a powerful platform for quantum computation as well.
There are some interesting studies on resonator qudits. For example,
Moon and  Girvin \cite{kerr} studied theoretically the parametric
down-conversion and squeezing of microwaves inside a transmission
line resonator, resorting to circuit QED in 2005.  In 2007,
Marquardt \cite{twin} presented an efficient scheme for the
generation of microwave photon pairs by parametric down-conversion
in a superconducting resonator coupled to a superconducting qubit.
In 2008, Hofheinz \emph{et al.} \cite{Hofheinz} demonstrated the
preparation of  pure Fock states with a microwave resonator,
resorting to a superconducting phase qubit. In the next year, they
synthesized arbitrary quantum states in a superconducting resonator
\cite{Hofheinz2}. In 2009, Rebi\'c \emph{et al.} \cite{kerr2}
introduced a scheme for generating  giant Kerr nonlinearities in
circuit QED. In 2010, Bergeal \emph{et al.} \cite{kerr1} proposed a
practical microwave device for achieving parametric amplification.
In this year, Johnson \emph{et al.} \cite{Johnson} demonstrated a
quantum nondemolition detection scheme that measures the number of
photons inside a high-quality-factor microwave cavity on a chip and
Strauch \emph{et al.} \cite{frederick} presented an effective method
to synthesize an arbitrary quantum state of two superconducting
resonators. In 2011, Mariantoni \emph{et al.} \cite{Mariantoni} used
a three-resonator circuit to shuffle one- and two-photon Fock states
between the three resonators and demonstrated qubit-mediated vacuum
Rabi swaps between two resonators. In addition, there are some
interesting works to generate the entanglement between the resonator
qudits \cite{noon,Wang,Han,Zakka,Tian,Friis,Siyuan,Frederick1}.

To realize the quantum computation based on resonator qudits, people
should construct the universal quantum gates on qudits. In 2007,
Schuster \emph{et al.} \cite{DISchuster} proposed the effect of the
number-state-dependent interaction between a superconducting qubit
and resonator qudits, which provides an interesting way to achieve
the state-selective qubit rotation. Based on this effect, Strauch
\cite{FWStrauch} presented an interesting scheme to construct the
c-phase gate on two superconducting resonator qudits in 2011. In his
work, each of two resonators (A and B) is coupled to an auxiliary
three-level transmon or phase qutrit (a and b), and each qutrit is
coupled to each other directly. The operation time of the c-phase
gate on two resonator qudits is 150 ns. In 2012, Wu \emph{et al.}
\cite{Wu} gave an effective scheme for the construction of the
c-phase gate on two resonators by using the number-state-dependent
interaction between a two-energy-level charge qubit and two
resonator-qudits for one-way quantum computation, and the operation
time of the c-phase gate was 125 ns.

In this paper, we give a microwave-photon quantum processor with two
resonators which are coupled to just one transmon qutrit which has
the  characteristics of a lesser anharmonicity energy level and a long
coherence time \cite{Koch}, and we construct an effective c-phase
gate on two resonator qudits, resorting to the combination of the
number-state-dependent interaction between the qutrit and one
resonator-qudit subsystem and the simple resonant operation between
the qutrit and another resonator-qudit subsystem. This different
physical mechanism  provides us  a faster way to achieve a
higher-fidelity c-phase gate on the two  resonator qudits without
increasing the difficulty of its implementation, compared with the
previous proposals \cite{FWStrauch,Wu}. The fidelity of our c-phase
gate is $99.51\%$ within the operation time of 93 ns. Moreover, our
device can be extended easily to construct the three-qudit cc-phase
gate  on  three resonator qudits, by using a resonator to complete
the simple resonant operation and the other two resonators to
achieve the number-state-dependent interaction on the transmon
qutrit, far different from the existing proposals. Its fidelity  is
$92.92\%$ within the operation time of 124.64 ns.

\begin{figure}[h]
\par
\begin{center}
\includegraphics[width=6.4cm,angle=0]{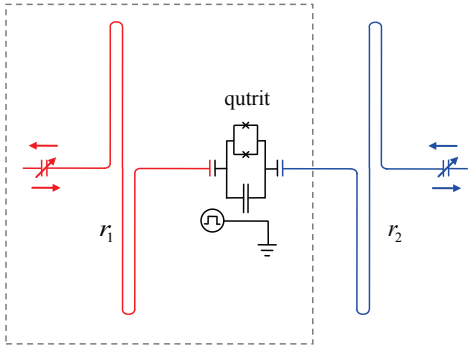}
\end{center}
\caption{(Color online) Schematic diagram  for our c-phase gate on
two microwave-photon-resonator qudits, by combination of the
number-state-dependent interaction between the transmon qutrit
and  the left resonator ($r_1$) and the simple resonant
operation between the transmon qutrit and  the right resonator ($r_2$). The
two resonators are capacitively coupled to the
qutrit whose transition frequency can be tuned by an
external flux.} \label{fig1}
\end{figure}

\section{Controlled-phase gate on two resonators in circuit QED}

Let us first consider a system composed of a perfect two-level
superconducting qubit $q$ and a resonator ($r_1$), whose
schematic diagram is the same as that shown in the dashed-line box
in Fig. \ref{fig1} (by replacing the three-energy-level qutrit with a two-energy-level qubit). The Hamiltonian for this system under the
rotating-wave approximation ($\hbar =1$) is
\begin{equation}  
H_1=\omega_{r_1}a^{+}a+\omega_q \sigma ^{+}\sigma ^{-}+g\left(a\sigma
^{+}+a^{+}\sigma ^{-}\right), \label{H1}
\end{equation}
where $\sigma^{+}=\vert 1\rangle \langle 0\vert $ and $a^{+}$ are
the creation operators of the superconducting qubit $q$ and the
resonator $r_1$, respectively. $g$ is the coupling strength between
the qubit and the resonator.  $\omega_{r_1}$ and $\omega_q$ are the
transition frequencies of the resonator $r_1$ and the qubit,
respectively.  In the dispersive regime ($\frac{g^{2}}{\Delta }$
$\leq 1$) in circuit QED, by making the unitary transformation $U=\exp
\left[ \frac{g}{\Delta }\left(a\sigma ^{+}-a^{+}\sigma
^{-}\right)\right]$, the Hamiltonian $ H_1$ becomes \cite{SQ}
\begin{equation} 
H'_1=UH_{1}U^{+}\approx \omega_{r_1}a^{+}a+\frac{1}{2}\left[\omega
_{q}+\frac{g^{2}}{ \Delta }(2a^{+}a+1)\right]\sigma _{z}.
\label{H11}
\end{equation}
The effect of the photon-number-dependent transition frequency of
the qubit  can be described as
\begin{equation} 
{\omega'}^{n}_{q}=\omega_{q}+\frac{g^{2}}{\Delta}(2n+1).
\label{stark}
\end{equation}
Here  $\Delta =\omega_{r_1}-\omega_{q}$. $n=a^{+}a$ is the photon
number in a resonator.  ${\omega'}^{n}_{q}$ is the changed
transition frequency of the qubit due to the different photon numbers
in the resonator.

In the dispersive strong regime ($\frac{g^{2}}{\Delta } \ll 1$) \cite{DISchuster}, the
photon-number-dependent transition frequency of the qubit is too
small to distinguish the different transition frequencies of the
qubit due to the different photon numbers in the resonator. By keeping
$\frac{g}{\Delta }$ to be a small value to make  Eq. (\ref{H11}) work well,
one can increase the coupling strength to make
the transition frequency of the
qubit depend on largely the photon number, shown in
Eq. (\ref{stark}). That is, if we apply a drive field with the
frequency equivalent to the transition frequency of the qubit when
$n=1$, and take the proper amplitude $ \left\vert \Omega \right\vert
\ll \frac{g^{2}}{\Delta }$ to suppress the error generated by
off-resonant transitions sufficiently, the field will flip the qubit
only if there is one microwave photon in the resonator. On the other
hand, if we apply a drive field with the frequency equivalent to the
transition frequency of the qubit when $n=0$, and take the proper
amplitude, the field will flip the qubit only if there is no
microwave photon in the resonator.  To describe this effect, we
consider a system with the resonator $r_{1}$ coupled to a practical
transmon qutrit \cite{Hoi}, whose Hamiltonian is (under the
rotating-wave approximation)
\begin{eqnarray} 
H_{2} &=& \sum_{l=g,e,f}E_{l}\left\vert l\right\rangle
_{q}\left\langle l\right\vert +\omega_{r_1}a_{1}^{+}a_{1}
+g_{r_1}^{g,e}(a_{1}^{+}\sigma _{g,e}^{-}+a_{1}\sigma
_{g,e}^{+})\nonumber\\
&&+g_{r_1}^{e,f}(a_{1}^{+}\sigma _{e,f}^{-}+a_{1}\sigma _{e,f}^{+}),
\label{selective}
\end{eqnarray}
where $\left\vert g\right\rangle_{q}$, $\left\vert
e\right\rangle_{q}$, and $\left\vert f\right\rangle_{q}$ are the
first three lower-energy levels of the qutrit. $\sigma _{g,e}^{+}$
and $\sigma _{e,f}^{+}$ are the creation operators for the
transitions $\left\vert g\right\rangle _{q}\rightarrow \left\vert
e\right\rangle _{q}$ and $\left\vert e\right\rangle _{q}\rightarrow
\left\vert f\right\rangle _{q}$\ of the qutrit $q$, respectively.
$a_{1}^{+}$ is the creation operator of the resonator $r_1$. The
energy for the level $l$ of $q$ is $E_{l}$, and $ \omega_{r_1}$ is
the transition frequency of $r_{1}$. $g_{1}^{g,e}$ and $g_{1}^{e,f}$
are the coupling strengths between these two transitions of $q$ and
$r_{1}$.

\begin{figure}[!h]
\par
\begin{center}
\includegraphics[width=7.8cm,angle=0]{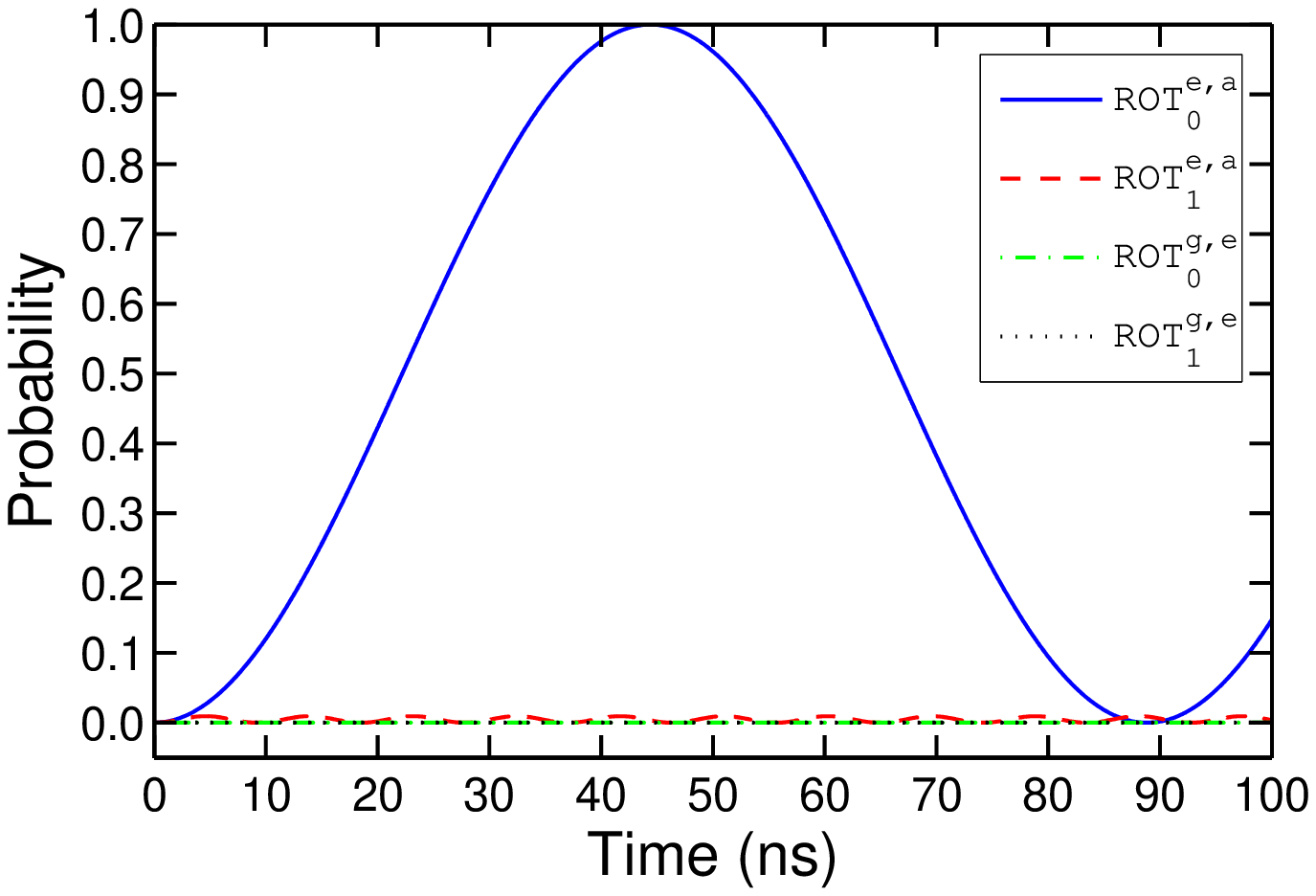}
\end{center}
\caption{(Color online) The expectation values of the probability
distributions of the quantum Rabi oscillations ROT$_0^{g,e}$,
ROT$_0^{e,f}$, ROT$_1^{g,e}$, and ROT$_1^{e,f}$. ROT$_0^{g,e}$ and
ROT$_0^{e,f}$ represent  the   oscillations of $(\left\vert
0\right\rangle _{r_1}\left\vert g\right\rangle
_{q})_{dress}\leftrightarrow (\left\vert 0\right\rangle
_{r_1}\left\vert e\right\rangle _{q})_{dress}$ and $(\left\vert
0\right\rangle _{r_1}\left\vert e\right\rangle
_{q})_{dress}\leftrightarrow (\left\vert 0\right\rangle
_{r_1}\left\vert f\right\rangle _{q})_{dress}$, respectively.
ROT$_1^{g,e}$ and ROT$_1^{e,f}$ represent the oscillations  of
$(\left\vert 1\right\rangle _{r_1}\left\vert g\right\rangle
_{q})_{dress}\leftrightarrow (\left\vert 1\right\rangle
_{r_1}\left\vert e\right\rangle _{q})_{dress}$ and $(\left\vert
1\right\rangle _{r_1}\left\vert e\right\rangle
_{q})_{dress}\leftrightarrow (\left\vert 1\right\rangle
_{r_1}\left\vert f\right\rangle _{q})_{dress}$, respectively}
\label{fig2}
\end{figure}

A microwave drive field $H_{d}=\Omega (\left\vert f\right\rangle
_{q}\left\langle e\right\vert e^{-i\omega _{d}t}+\left\vert
e\right\rangle _{q}\left\langle f\right\vert e^{i\omega _{d}t})$
with a proper amplitude $\Omega$ is applied to interact with the
qutrit, and here the frequency $\omega _{d}$ is chosen to be
equivalent to the transition frequency ($\left\vert
e\right\rangle_{q}\leftrightarrow\left\vert f\right\rangle _{q}$) of
the qutrit $q$ when there is no microwave photon in the resonator.
Due to the realistic quantum Rabi oscillation (ROT) occurring
between the dress states of the system \cite{Strauch}, we simulate
the expectation value of ROT$_0^{e,f}$ $(\left\vert 0\right\rangle
_{r_1}\left\vert e\right\rangle _{q})_{dress}\leftrightarrow
(\left\vert 0\right\rangle _{r_1}\left\vert f\right\rangle
_{q})_{dress}$ and ROT$_1^{e,f}$ $(\left\vert 1\right\rangle
_{r_1}\left\vert e\right\rangle _{q})_{dress}\leftrightarrow
(\left\vert 1\right\rangle _{r_1}\left\vert f\right\rangle
_{q})_{dress}$, shown in Fig. \ref{fig2}.  The transition
frequencies of $\left\vert g\right\rangle _{q}\leftrightarrow
\left\vert e\right\rangle _{q}$ and $ \left\vert e\right\rangle
_{q}\leftrightarrow \left\vert f\right\rangle _{q}$ of the qutrit
are chosen to be $\omega_{g,e}/(2\pi )=E_e-E_g=8.7GHz$ and $\omega
_{e,f}/(2\pi )=E_f-E_e=8.0GHz$, respectively. $\omega_{r_1}/(2\pi
)=7.5GHz$.  The coupling strengths between two transitions of the
qutrit and $r_{1}$ are taken in convenience with $g_{1}^{g,e}/(2\pi
)=g_{1}^{e,f}/(2\pi )=0.2GHz$. The frequency and amplitude of the
drive field are $\omega_{d}/(2\pi )=8.043GHz$ and $ \Omega
=0.0115GHz$, respectively.

As shown in Fig. \ref{fig2}, the maximal probability of
ROT$_{0}^{e,f}$ can reach $100\%$.  After a period of
ROT$_{0}^{e,f}$, a $\pi $ phase shift can be generated in the state
$ (\left\vert 0\right\rangle _{1}\left\vert e\right\rangle
_{q})_{dress}$, and ROT$_{1}^{e,f}$ and the other oscillations take
place with a very small probability, which indicates that the final
state of the system composed of $r_{1}$ and $q$ becomes
\begin{eqnarray} 
\left\vert \phi _{f}\right\rangle &=& \frac{1}{2}[(\left\vert
0\right\rangle _{1}\left\vert g\right\rangle
_{q})_{dress}-(\left\vert 0\right\rangle
_{1}\left\vert e\right\rangle _{q})_{dress} \nonumber\\
&&+(\left\vert 1\right\rangle _{1}\left\vert g\right\rangle
_{q})_{dress}+(\left\vert 1\right\rangle _{1}\left\vert
e\right\rangle _{q})_{dress}]
\end{eqnarray}
after the state-selective qubit rotation if  the initial state of
the system is
\begin{eqnarray} 
\left\vert \phi _{0}\right\rangle &=& \frac{1}{2}[(\left\vert 0\right\rangle
_{1}\left\vert g\right\rangle _{q})_{dress}+(\left\vert 0\right\rangle
_{1}\left\vert e\right\rangle _{q})
_{dress} \nonumber\\
&&+(\left\vert 1\right\rangle _{1}\left\vert g\right\rangle
_{q})_{dress}+(\left\vert 1\right\rangle _{1}\left\vert
e\right\rangle _{q})_{dress}].
\end{eqnarray}
This is just the outcome of a hybrid c-phase gate on $r_1$ and $q$
by using  $r_1$  as the control qubit and  $q$ as the target qubit.
At the beginning and the end of the algorithm, one can turn on and off
the coupling between $r_1$ and $q$ to evolve the dress states into
the computational states \cite{Strauch}.  On one hand, one can tune
the transition frequency of the resonator or the qutrit to make them
resonate or largely detune with each other, in order to turn on or
off the interaction between $q$ and a non-computational resonator.
On the other hand, one can tune on or off the coupling  between the
qutrit and the noncomputational resonator \cite{Allman,Bialczak}.

The principle of our c-phase gate on two resonator qudits based on
the state-selective qubit rotation is shown in Fig. \ref{fig1}. The
matrix representation of the c-phase gate can be written as
\begin{eqnarray} 
\left(
\begin{array}{cccc}
1 & 0 & 0 & 0 \\
0 & 1 & 0 & 0 \\
0 & 0 & 1 & 0 \\
0 & 0 & 0 & -1
\end{array}
\right)
\label{matrix}
\end{eqnarray}
in the basis of a two-resonator-qudit system {$\left\vert
0\right\rangle _{1}\left\vert 0\right\rangle _{2}$, $\left\vert
0\right\rangle _{1}\left\vert 1\right\rangle _{2}$, $\left\vert
1\right\rangle _{1}\left\vert 0\right\rangle _{2}$, $\left\vert
1\right\rangle _{1}\left\vert 1\right\rangle _{2}$\}. The
Hamiltonian of the system composed of the resonators $r_1$, $r_2$,
and $q$ can be written as (under the rotating-wave approximation)
\begin{eqnarray}  
H_2 &=&\sum\limits_{l=g,e,f}E_{l}\left\vert l\right\rangle
_{q}\left\langle l\right\vert   + \sum_{i=1,2}[\omega
_{r_i}a_{i}^{+}a_{i}+g_{i}^{g,e}(a_{i}^{+}\sigma
_{g,e}^{-}+a_{i}\sigma _{g,e}^{+})\nonumber\\
&&+g_{i}^{e,f}(a_{i}^{+}\sigma _{e,f}^{-}+a_{i}\sigma _{e,f}^{+})].
\label{allhamiltonian}
\end{eqnarray}
Suppose that the initial state of the system is
\begin{equation} 
\left\vert \psi _{0}\right\rangle = \frac{1}{2}(\left\vert 0\right\rangle
_{1}\left\vert 0\right\rangle _{2}+\left\vert
0\right\rangle _{1}\left\vert 1\right\rangle
_{2}
+\left\vert 1\right\rangle _{1}\left\vert
0\right\rangle _{2}+\left\vert 1\right\rangle _{1}
\left\vert 1\right\rangle _{2})\otimes\left\vert g\right\rangle _{q}.
\label{cphase0}
\end{equation}
Our c-phase gate on the two resonators can be accomplished with
three steps as follows.

\begin{figure*}[tpb] 
\begin{center}
\includegraphics[width=6.0cm,angle=0]{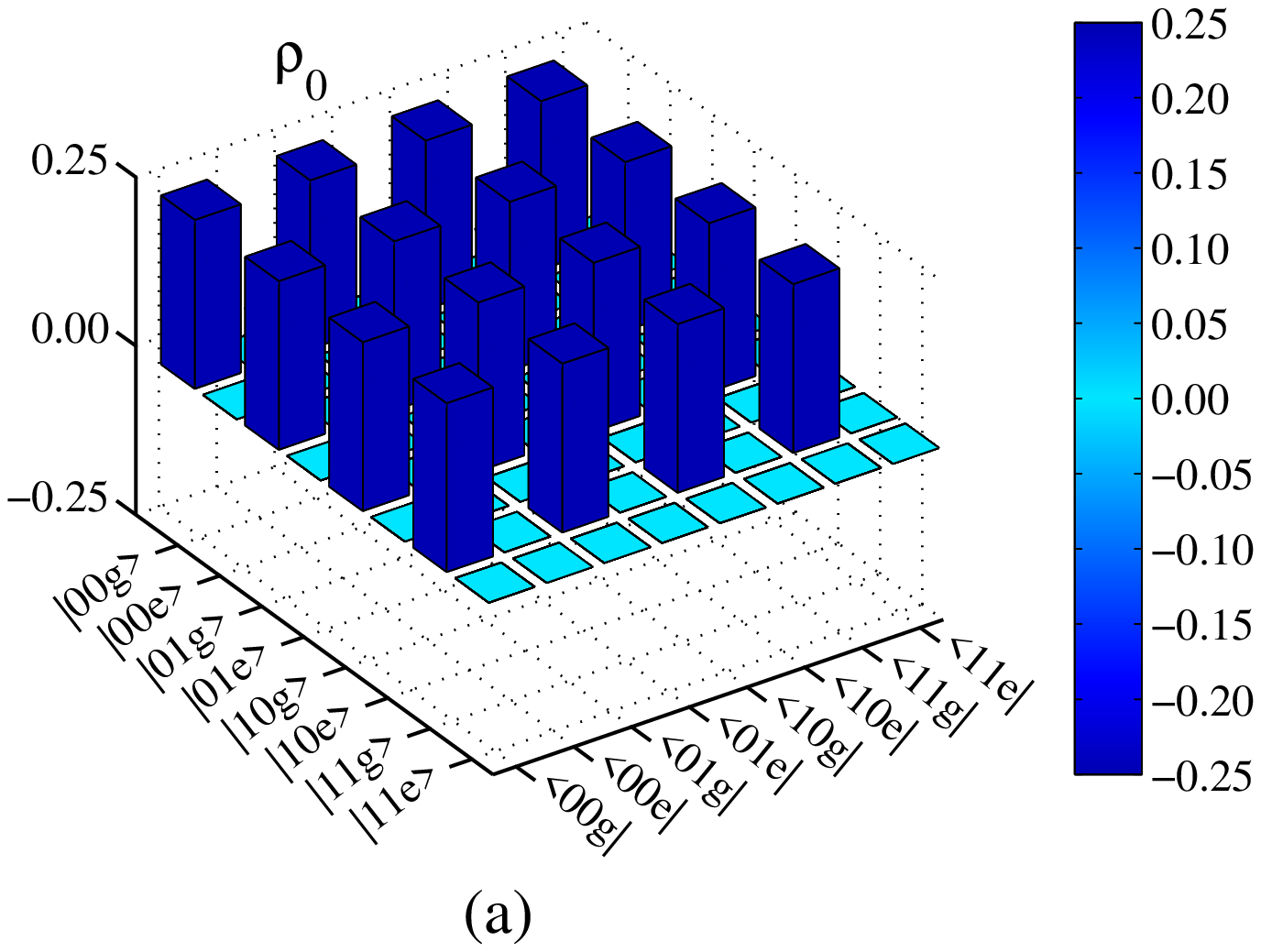}$\;\;\;\;\;\;\;\;\;\; $
\includegraphics[width=4.8cm,angle=0]{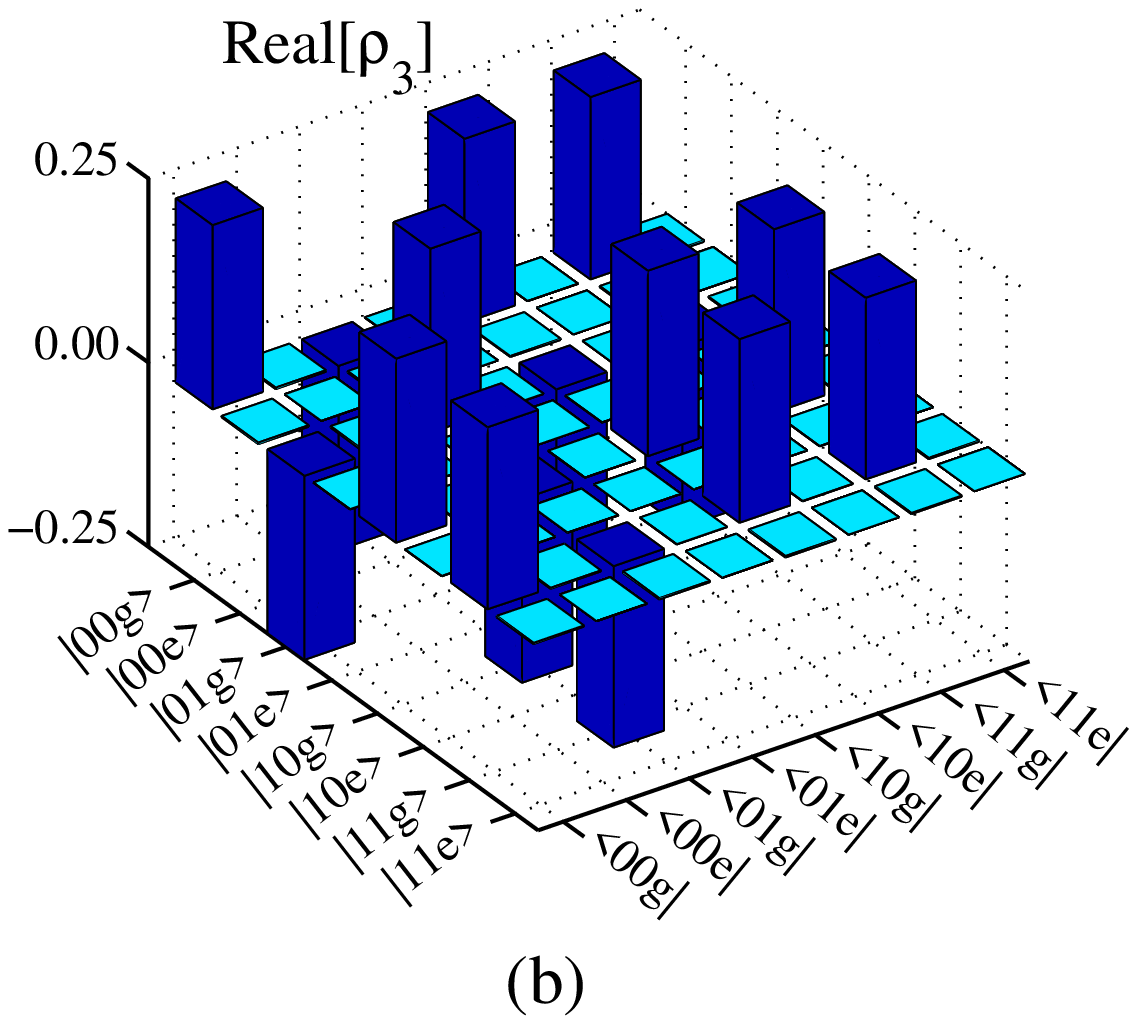}$\;\;\;\;\;\;\;\; $
\includegraphics[width=4.8cm,angle=0]{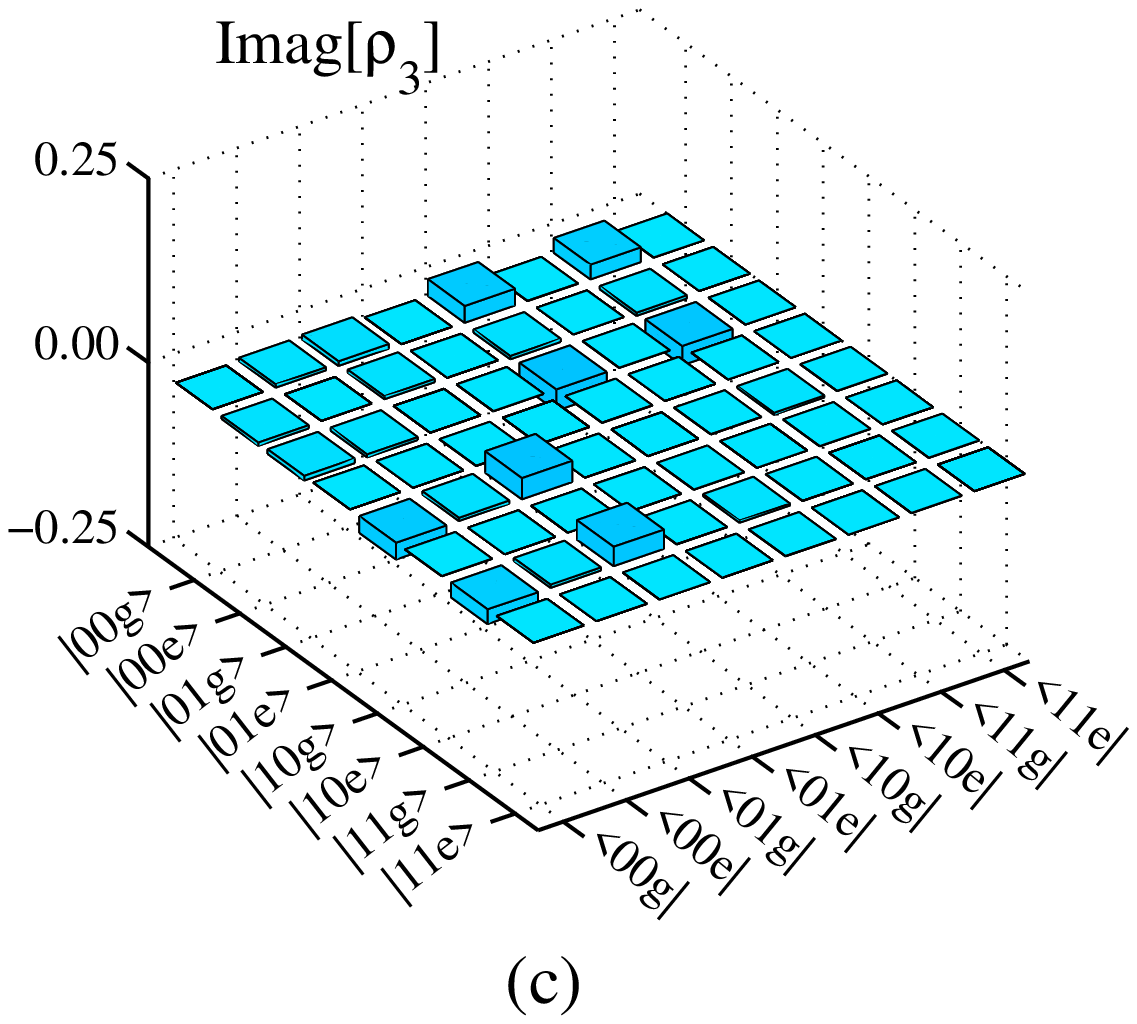}
\end{center}
\caption{(Color online) (a) The density operator ($\rho_0$) of the
initial state $\left\vert \psi _{0}\right\rangle$ of the system
composed of the two resonators and the qutrit in our c-phase gate. Panels
(b) and (c) are the real part (Real[$\rho_3$]) and the imaginary part
(Imag[$\rho_3$]) of the final state $\left\vert \psi
_{f}\right\rangle$ of the system, respectively.} \label{fig3}
\end{figure*}

First, by resonating $r_{2}$ and  $q$ with
$g^{g,e}_{2}t=\frac{\pi}{2}$, and turning off the interaction
between $r_{1}$ and $q$, the system evolves from the initial state
$\left\vert \psi _{0}\right\rangle$ to the state
\begin{equation} 
\left\vert \psi _{1}\right\rangle = \frac{1}{2}(\left\vert 0\right\rangle
_{1}\left\vert g\right\rangle _{q}-i\left\vert
0\right\rangle _{1}\left\vert e\right\rangle _{q}\nonumber\\
+\left\vert 1\right\rangle _{1}\left\vert g\right\rangle _{q}
-i\left\vert 1\right\rangle _{1}\left\vert e\right\rangle
_{q})\otimes\left\vert 0\right\rangle _{2}.
\label{cphase1}
\end{equation}

Second, by turning on the coupling between $r_{1}$ and $q$, and
turning off the coupling between $r_{2}$ and $q$, the state of the
system becomes
\begin{eqnarray} 
\left\vert \psi _{1}^{\prime }\right\rangle &=&
\frac{1}{2}[(\left\vert 0\right\rangle _{1}\left\vert g\right\rangle
_{q})_{dress}-i(\left\vert 0\right\rangle _{1}\left\vert
e\right\rangle _{q})_{dress}  \nonumber \\
&& +(\left\vert 1\right\rangle _{1}\left\vert g\right\rangle
_{q})_{dress}-i(\left\vert 1\right\rangle _{1}\left\vert
e\right\rangle _{q})_{dress}]\otimes\left\vert 0\right\rangle _{2}.
\label{cphase21}
\end{eqnarray}
By applying  a drive field $H_{d}=\Omega (\left\vert f\right\rangle
_{q}\left\langle e\right\vert e^{-i\omega _{d}t}+\left\vert
e\right\rangle _{q}\left\langle f\right\vert e^{i\omega _{d}t})$ with the
frequency equivalent to the transition frequency ($\left\vert
e\right\rangle _{q}\leftrightarrow \left\vert f\right\rangle _{q}$)
of the qutrit when there is no  microwave photon in the resonator
$r_1$, after an operation time of $\Omega t=\pi$, the
state of the system is changed to be
\begin{eqnarray} 
\left\vert \psi _{2}\right\rangle &=& \frac{1}{2}[(\left\vert
0\right\rangle _{1}\left\vert g\right\rangle
_{q})_{dress}+i(\left\vert 0\right\rangle _{1}\left\vert
e\right\rangle _{q})_{dress}  \nonumber \\
&& +(\left\vert 1\right\rangle _{1}\left\vert g\right\rangle
_{q})_{dress}-i(\left\vert 1\right\rangle _{1}\left\vert
e\right\rangle _{q})_{dress}]\otimes\left\vert 0\right\rangle _{2}.
\label{cphase22}
\end{eqnarray}

Third, by turning off the coupling between {$r_{1}$ and $q$, the
state of the system  evolves from $\left\vert \psi
_{2}\right\rangle$ into
\begin{equation} 
\left\vert \psi _{2}^{\prime }\right\rangle =\frac{1}{2}(\left\vert
0\right\rangle _{1}\left\vert g\right\rangle
_{q}+i\left\vert 0\right\rangle _{1}\left\vert
e\right\rangle _{q}+\left\vert 1\right\rangle _{1}\left\vert g\right\rangle _{q}-i\left\vert 1\right\rangle _{1}
\left\vert e\right\rangle _{q})\otimes\left\vert 0\right\rangle _{2}.
\label{cphase31}
\end{equation}

By resonating $r_{2}$ and  $q$ with $g^{g,e}_{2}t=\frac{\pi }{2}$
again, and turning off the interaction between $r_{1}$ and $q$, the
state of the system becomes
\begin{eqnarray} 
\left\vert \psi _{f}\right\rangle =\frac{1}{2}(\left\vert
0\right\rangle _{1}\left\vert 0\right\rangle _{2}\!+\!\left\vert
0\right\rangle _{1}\left\vert 1\right\rangle _{2} \!+\!\left\vert
1\right\rangle _{1}\left\vert 0\right\rangle _{2} \!-\!\left\vert
1\right\rangle _{1}\left\vert 1\right\rangle
_{2})\!\otimes\!\left\vert g\right\rangle _{q}. \label{finalstate}
\end{eqnarray}
This is just the result of the c-phase gate on $r_{1}$ and $r_{2}$
by using  $r_{1}$  as  the control qubit and  $r_{2}$ as  the target
qubit.

The reduced density operators of  the two-resonator system in the
initial state $\vert \psi_0\rangle$ in Eq. (\ref{cphase0}) and the final state $\vert \psi_f\rangle$ in  Eq.
(\ref{finalstate}) are shown in Fig. \ref{fig3}. One can see that
the fidelity of our c-phase gate on two microwave-photon qudits is
about $99.51\%$ within about 93 ns. Here the fidelity is defined as
$F=Tr(\left\vert \sqrt{\rho_{f}}\,\rho_{ideal}\sqrt{\rho_{f}}
\right\vert )$ \cite{Haack}. $\rho _{f}$ is the density operator of the
final state of the two-microwave-photon-qudit system  $\left\vert
\psi _{f}\right\rangle $ and $\rho _{ideal}$ is the density operator
of the final state of the system after an ideal c-phase gate
operation is performed with the initial state $ {\left\vert \psi
_{0}\right\rangle }$.

\begin{figure}[!tpb]
\par
\begin{center}
\includegraphics[width=8.0cm,angle=0]{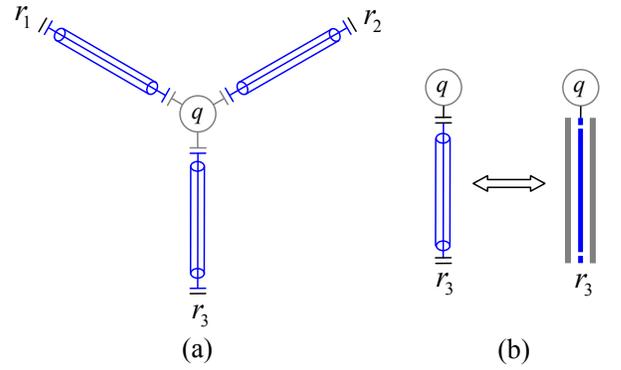}
\end{center}
\caption{(Color online) (a) The schematic diagram for our
cc-phase gate on a  three-qudit microwave-photon
system. (b) The schematic diagram for the coupling between a
transmon qutrit and a microwave-photon resonator. Here $q$
represents a transmon qutrit. $r_1$, $r_2$, and $r_3$ are
three microwave-photon resonators which have the same structure as
those shown in Fig.\ref{fig1}.} \label{fig4}
\end{figure}

\section{Controlled-controlled-phase gate  on three resonators}

The principle of our cc-phase gate on a three-resonator system is
shown in Fig.\ref{fig4}. Here the resonator $r_1$ has the same role
as $r_2$ and they both are used to provide the  effect of the
photon-number-dependent transition frequency of the $\Xi$-type
three-level qutrit  to accomplish the state-selective qubit
rotation, different from the resonator $r_3$.  The
photon-number-dependent transition frequency between $ \left\vert
e\right\rangle_{q}$ and  $\left\vert f\right\rangle _{q}$ of
the qutrit can be written as \cite{frederick,Strauch,Wu}
\begin{equation} 
{\omega'}^{n_{1},\,n_{2}}_{\,e,\,f}\approx \omega
_{e,\,f}+\frac{(g_{1}^{e,\,f})^{2}}{\omega
_{e,\,f}-\omega_{r_1}}(2n_{1}+1)+\frac{(g_{2}^{e,\,f})^{2}}{\omega
_{e,\,f}-\omega_{r_2}} (2n_{2}+1).  \label{ccphasefrequency}
\end{equation}
Here $n_1$ and $n_2$ are the photon numbers in the resonators $r_1$
and $r_2$, respectively. The photon-number-dependent transition
frequency of $q$ depends on the relationship of the photon numbers
in two resonators. That is,  one  can afford a drive field with the
frequency equivalent to the changed transition frequency of the
qutrit to achieve the state-selective qubit rotation  with different
relations between $n_{1}$ and $n_{2}$.

Suppose that $\frac{3(g_{1}^{e,\,f})^{2}}{ \omega
_{e,f}-\omega_{r_1}} =\frac{(g_{2}^{e,\,f})^{2}}{\omega
_{e,f}-\omega_{r_2}}$, one can obtain the relation
${\omega'}^{n_{1},\,n_{2}}_{\,e,\,f}=\omega _{e,\,f}+
\frac{(g_{1}^{e,\,f})^{2}}{\omega_{e,\,f}-\omega_{r_1}}(N+4)$,
where $N=2n_{1}+6n_{2}$. The transition frequency of $q$ can be
divided into four groups, according to the photon-number relations
between $r_{1}$ and $r_{2}$. That is,  $\left\vert 0\right\rangle
_{1}\left\vert 0\right\rangle _{2}$, $\left\vert 1\right\rangle
_{1}\left\vert 0\right\rangle _{2}$, $\left\vert 0\right\rangle
_{1}\left\vert 1\right\rangle _{2}$, and $\left\vert 1\right\rangle
_{1}\left\vert 1\right\rangle _{2}$ with $N=0,2,6$, and $8$,
respectively. Considering $N=8$ , a drive field   with the frequency
 $\omega_{d}=\omega
_{e,f}+\frac{12(g_{1}^{e,\,f})^{2}}{\omega_{e,f}-\omega_{r_1}}$
can flip the qutrit between $\left\vert e\right\rangle_{q}$ and
$\left\vert f\right\rangle_{q}$ only if there is one microwave
photon in each of the two resonators $r_1$ and $r_2$.

If we take the initial state of the hybrid system composed of
$r_1$, $r_2$, and $q$ as
\begin{eqnarray} 
\left\vert \Phi _{0}\right\rangle  &=&\frac{1}{2\sqrt{2}}[(\left\vert
0\right\rangle _{1}\left\vert 0\right\rangle _{2}\left\vert g\right\rangle
_{q})_{dress}+(\left\vert 0\right\rangle _{1}\left\vert 0\right\rangle
_{2}\left\vert e\right\rangle _{q})_{dress} \nonumber \\
&&+(\left\vert 0\right\rangle
_{1}\left\vert 1\right\rangle _{2}\left\vert g\right\rangle
_{q})_{dress}+(\left\vert 0\right\rangle _{1}\left\vert 1\right\rangle
_{2}\left\vert e\right\rangle _{q})_{dress} \nonumber \\
&&+(\left\vert 1\right\rangle _{1}\left\vert 0\right\rangle _{2}\left\vert
g\right\rangle _{q})_{dress}+(\left\vert 1\right\rangle _{1}\left\vert
0\right\rangle _{2}\left\vert e\right\rangle _{q})_{dress} \nonumber \\
&&+(\left\vert
1\right\rangle _{1}\left\vert 1\right\rangle _{2}\left\vert g\right\rangle
_{q})_{dress}+(\left\vert 1\right\rangle _{1}\left\vert 1\right\rangle
_{2}\left\vert e\right\rangle _{q})_{dress}], \label{2r0}
\end{eqnarray}
applying a drive field to complete a  state-selective qubit rotation
operation on the transition between $\left\vert e\right\rangle
_{q}$ and  $\left\vert f\right\rangle _{q}$ when there is
one microwave photon in each of the two resonators $r_{1}$ and
$r_{2}$,   the state of the hybrid system becomes
\begin{eqnarray} 
\left\vert \Phi _{f}\right\rangle  &=&\frac{1}{2\sqrt{2}}[(\left\vert
0\right\rangle _{1}\left\vert 0\right\rangle _{2}\left\vert g\right\rangle
_{q})_{dress}+(\left\vert 0\right\rangle _{1}\left\vert 0\right\rangle
_{2}\left\vert e\right\rangle _{q})_{dress} \nonumber \\
&&+(\left\vert 0\right\rangle
_{1}\left\vert 1\right\rangle _{2}\left\vert g\right\rangle
_{q})_{dress}+(\left\vert 0\right\rangle _{1}\left\vert 1\right\rangle
_{2}\left\vert e\right\rangle _{q})_{dress} \nonumber \\
&&+(\left\vert 1\right\rangle _{1}\left\vert 0\right\rangle _{2}\left\vert
g\right\rangle _{q})_{dress}+(\left\vert 1\right\rangle _{1}\left\vert
0\right\rangle _{2}\left\vert e\right\rangle _{q})_{dress} \nonumber \\
&&+(\left\vert 1\right\rangle _{1}\left\vert 1\right\rangle
_{2}\left\vert g\right\rangle _{q})_{dress}-(\left\vert
1\right\rangle _{1}\left\vert 1\right\rangle _{2}\left\vert
e\right\rangle _{q})_{dress}] \label{2r1}
\end{eqnarray}
after the operation time $t=\frac{\pi }{ \Omega}$. This is
just the result of a hybrid cc-phase gate on the system composed of
$r_{1}$, $r_{2}$, and $q$, by using $r_{1}$ and $r_{2}$ as the
control qudits and $q$ as the target qubit.

With the hybrid cc-phase gate above,  we can construct the cc-phase
gate on three resonator qudits, shown in Fig.\ref{fig4}. The
Hamiltonian of the hybrid system composed of the three resonators
$r_{1}$, $r_{2}$, and $r_{3}$ and the transmon qutrit $q$ is
\begin{eqnarray}  
H_3 &=& \sum\limits_{l=g,e,f}E_{l}\left\vert l\right\rangle
_{q}\left\langle l \,\right\vert   + \sum_{i=1,2,3}[\omega
_{i}^{r}a_{i}^{+}a_{i}+g_{i}^{g,e}(a_{i}^{+}\sigma
_{g,e}^{-}+a_{i}\sigma _{g,e}^{+})\nonumber\\
&&+g_{i}^{e,f}(a_{i}^{+}\sigma _{e,f}^{-}+a_{i}\sigma _{e,f}^{+})].
\label{allhamiltonian}
\end{eqnarray}
Suppose that the initial state of the system is
\begin{eqnarray}  
\left\vert \Psi _{0}\right\rangle  &=&\frac{1}{2\sqrt{2}}(\left\vert
0\right\rangle _{1}\left\vert 0\right\rangle _{2}\left\vert 0\right\rangle
_{3}+\left\vert 0\right\rangle _{1}\left\vert 0\right\rangle _{2}\left\vert
1\right\rangle _{3} \nonumber\\
&&+\left\vert 0\right\rangle _{1}\left\vert 1\right\rangle
_{2}\left\vert 0\right\rangle _{3}+\left\vert 0\right\rangle _{1}\left\vert
1\right\rangle _{2}\left\vert 1\right\rangle _{3} \nonumber\\
&&+\left\vert 1\right\rangle _{1}\left\vert 0\right\rangle _{2}\left\vert
0\right\rangle _{3}+\left\vert 1\right\rangle _{1}\left\vert 0\right\rangle
_{2}\left\vert 1\right\rangle _{3} \nonumber\\
&&+\left\vert 1\right\rangle _{1}\left\vert 1\right\rangle
_{2}\left\vert 0\right\rangle _{3}+\left\vert 1\right\rangle
_{1}\left\vert 1\right\rangle _{2}\left\vert 1\right\rangle
_{3})\otimes \left\vert g\right\rangle _{q}. \label{ccphase0}
\end{eqnarray}
The cc-phase gate can be achieved with three steps as follows.

First, we turn off the interaction between the two resonators $r_{1}
r_{2}$ and $q$, and then resonate $r_3$ and the qutrit  $q$ in the
transition between $\left\vert g\right\rangle _{q}$ and $\left\vert
e\right\rangle_{q}$. After the operation time  $t=\frac{\pi
}{2g_3^{g,e}}$, the state of the system  evolves from $\left\vert \Psi
_{0}\right\rangle$ into
\begin{eqnarray} 
\left\vert \Psi _{1}\right\rangle  &=&\frac{1}{2\sqrt{2}}(\left\vert
0\right\rangle _{1}\left\vert 0\right\rangle _{2}\left\vert g\right\rangle
_{q}+i\left\vert 0\right\rangle _{1}\left\vert 0\right\rangle _{2}\left\vert
e\right\rangle _{q} \nonumber\\
&&+\left\vert 0\right\rangle _{1}\left\vert 1\right\rangle
_{2}\left\vert g\right\rangle _{q}+i\left\vert 0\right\rangle _{1}\left\vert
1\right\rangle _{2}\left\vert e\right\rangle _{q} \nonumber\\
&&+\left\vert 1\right\rangle _{1}\left\vert 0\right\rangle _{2}\left\vert
g\right\rangle _{q}+i\left\vert 1\right\rangle _{1}\left\vert 0\right\rangle
_{2}\left\vert e\right\rangle _{q} \nonumber\\
&&+\left\vert 1\right\rangle _{1}\left\vert
1\right\rangle _{2}\left\vert g\right\rangle _{q}+i\left\vert 1\right\rangle
_{1}\left\vert 1\right\rangle _{2}\left\vert e\right\rangle _{q})\otimes
\left\vert 0\right\rangle _{3}.
\label{ccphase1}
\end{eqnarray}

Second, we  turn off the interaction between $r_3$ and $q$, and turn
on the interactions between $r_1$ and $q$ and between $r_2$ and $q$. The
state $\left\vert \Psi _{1}\right\rangle$  is changed to be
\begin{eqnarray}
\left\vert \Psi' _{1}\right\rangle  &=&\frac{1}{2\sqrt{2}}
[(\left\vert 0\right\rangle _{1}\left\vert 0\right\rangle
_{2}\left\vert g\right\rangle _{q})_{dress}+i(\left\vert
0\right\rangle _{1}\left\vert
0\right\rangle _{2}\left\vert e\right\rangle _{q})_{dress} \nonumber\\
&&+(\left\vert 0\right\rangle _{1}\left\vert 1\right\rangle _{2}\left\vert
g\right\rangle _{q})_{dress}+i(\left\vert 0\right\rangle _{1}\left\vert
1\right\rangle _{2}\left\vert e\right\rangle _{q})_{dress} \nonumber\\
&&+(\left\vert 1\right\rangle _{1}\left\vert 0\right\rangle _{2}\left\vert
g\right\rangle _{q})_{dress}+i(\left\vert 1\right\rangle _{1}\left\vert
0\right\rangle _{2}\left\vert e\right\rangle _{q})_{dress} \nonumber\\
&&+(\left\vert 1 \!\right\rangle _{1}\!\left\vert 1 \!\right\rangle
_{2} \!\left\vert g\right\rangle _{q})_{dress}\!+\!i(\left\vert 1
\!\right\rangle _{1}\left\vert 1 \!\right\rangle _{2}\left\vert
e\right\rangle _{q})_{dress}] \!\otimes \! \left\vert 0\right\rangle
_{3}.\;\;\;\;\;\;\; \label{ccphase21}
\end{eqnarray}
By taking the hybrid cc-phase gate on $r_{1}$, $r_{2}$, and $q$, we
can get
\begin{eqnarray}
\left\vert \Psi _{2}\right\rangle  &=&\frac{1}{2\sqrt{2}}[(\left\vert
0\right\rangle _{1}\left\vert 0\right\rangle _{2}\left\vert g\right\rangle
_{q})_{dress}+i(\left\vert 0\right\rangle _{1}\left\vert 0\right\rangle
_{2}\left\vert e\right\rangle _{q})_{dress} \nonumber\\
&&+(\left\vert 0\right\rangle _{1}\left\vert 1\right\rangle _{2}\left\vert
g\right\rangle _{q})_{dress}+i(\left\vert 0\right\rangle _{1}\left\vert
1\right\rangle _{2}\left\vert e\right\rangle _{q})_{dress} \nonumber\\
&&+(\left\vert 1\right\rangle _{1}\left\vert 0\right\rangle _{2}\left\vert
g\right\rangle _{q})_{dress}+i(\left\vert 1\right\rangle _{1}\left\vert
0\right\rangle _{2}\left\vert e\right\rangle _{q})_{dress} \nonumber\\
&&+(\left\vert 1 \!\right\rangle _{1} \!\left\vert 1 \!\right\rangle
_{2} \!\left\vert g\right\rangle _{q})_{dress}\!-\!i(\left\vert 1
\!\right\rangle _{1} \!\left\vert 1 \!\right\rangle _{2}
\!\left\vert e\right\rangle _{q})_{dress}]\!\otimes\! \left\vert
0\right\rangle _{3}.\;\;\;\;\;\;\; \label{ccphase22}
\end{eqnarray}

Third, we turn off the coupling between $r_{1} r_{2}$ and $q$, the
state of the system   becomes
\begin{eqnarray}
\left\vert \Psi _{2}^{\prime }\right\rangle  &=&\frac{1}{2\sqrt{2}}%
[\left\vert 0\right\rangle _{1}\left\vert 0\right\rangle _{2}\left\vert
g\right\rangle _{q}+i\left\vert 0\right\rangle _{1}\left\vert 0\right\rangle
_{2}\left\vert e\right\rangle _{q} \nonumber\\
&&+\left\vert 0\right\rangle _{1}\left\vert 1\right\rangle _{2}\left\vert
g\right\rangle _{q}+i\left\vert 0\right\rangle _{1}\left\vert 1\right\rangle
_{2}\left\vert e\right\rangle _{q} \nonumber\\
&&+\left\vert 1\right\rangle _{1}\left\vert 0\right\rangle _{2}\left\vert
g\right\rangle _{q}+i\left\vert 1\right\rangle _{1}\left\vert 0\right\rangle
_{2}\left\vert e\right\rangle _{q} \nonumber\\
&&+\left\vert 1\right\rangle _{1}\left\vert 1\right\rangle
_{2}\left\vert g\right\rangle _{q}-i\left\vert 1\right\rangle
_{1}\left\vert 1\right\rangle _{2}\left\vert e\right\rangle
_{q}]\otimes \left\vert 0\right\rangle _{3}. \label{ccphase3}
\end{eqnarray}
By resonating $r_3$ and $q$, we can get the final state of the
system as
\begin{eqnarray}
\left\vert \Psi _{f}\right\rangle  &=&\frac{1}{2\sqrt{2}}[\left\vert
0\right\rangle _{1}\left\vert 0\right\rangle _{2}\left\vert 0\right\rangle
_{3}+\left\vert 0\right\rangle _{1}\left\vert 0\right\rangle _{2}\left\vert
1\right\rangle _{3} \nonumber\\
&&+\left\vert 0\right\rangle _{1}\left\vert 1\right\rangle _{2}\left\vert
0\right\rangle _{3}+\left\vert 0\right\rangle _{1}\left\vert 1\right\rangle
_{2}\left\vert 1\right\rangle _{3} \nonumber\\
&&+\left\vert 1\right\rangle _{1}\left\vert 0\right\rangle _{2}\left\vert
0\right\rangle _{3}+\left\vert 1\right\rangle _{1}\left\vert 0\right\rangle
_{2}\left\vert 1\right\rangle _{3} \nonumber\\
&&+\left\vert 1\right\rangle _{1}\left\vert 1\right\rangle
_{2}\left\vert 0\right\rangle _{3}-\left\vert 1\right\rangle
_{1}\left\vert 1\right\rangle _{2}\left\vert 1\right\rangle
_{3}]\otimes \left\vert g\right\rangle _{q}. \;\;\;\;\;\;\;\;
\label{ccphasefinal}
\end{eqnarray}
This is just the outcome of the cc-phase gate operation on $r_{1}$,
$r_{2}$, and $r_3$, by using $r_{1}$ and $r_{2}$ as the control
qudits and $r_3$ as the target qudit.

\begin{figure}[tpb]
\par
\begin{center}
\includegraphics[width=7.4cm,angle=0]{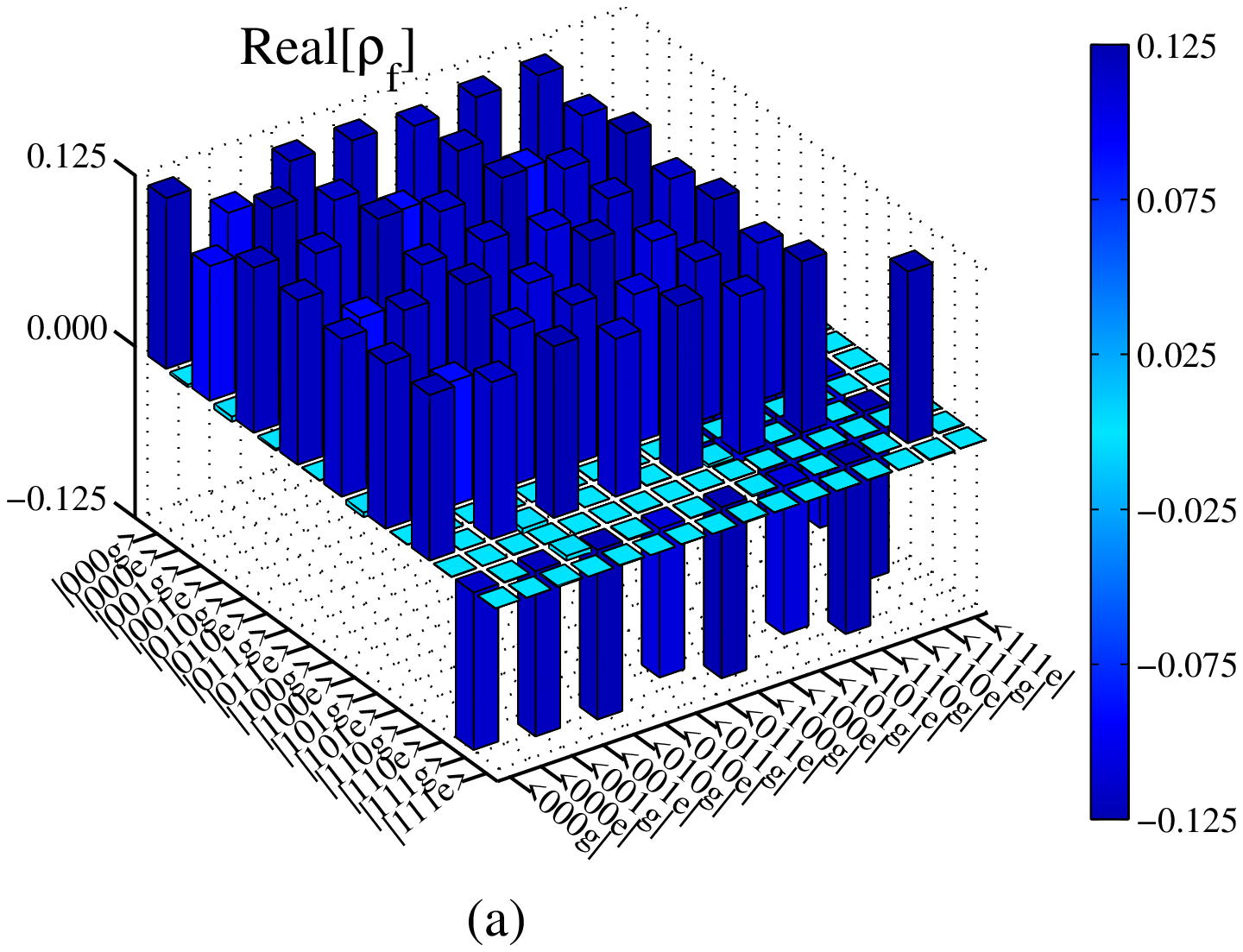}\\
\bigskip
\includegraphics[width=6.4cm,angle=0]{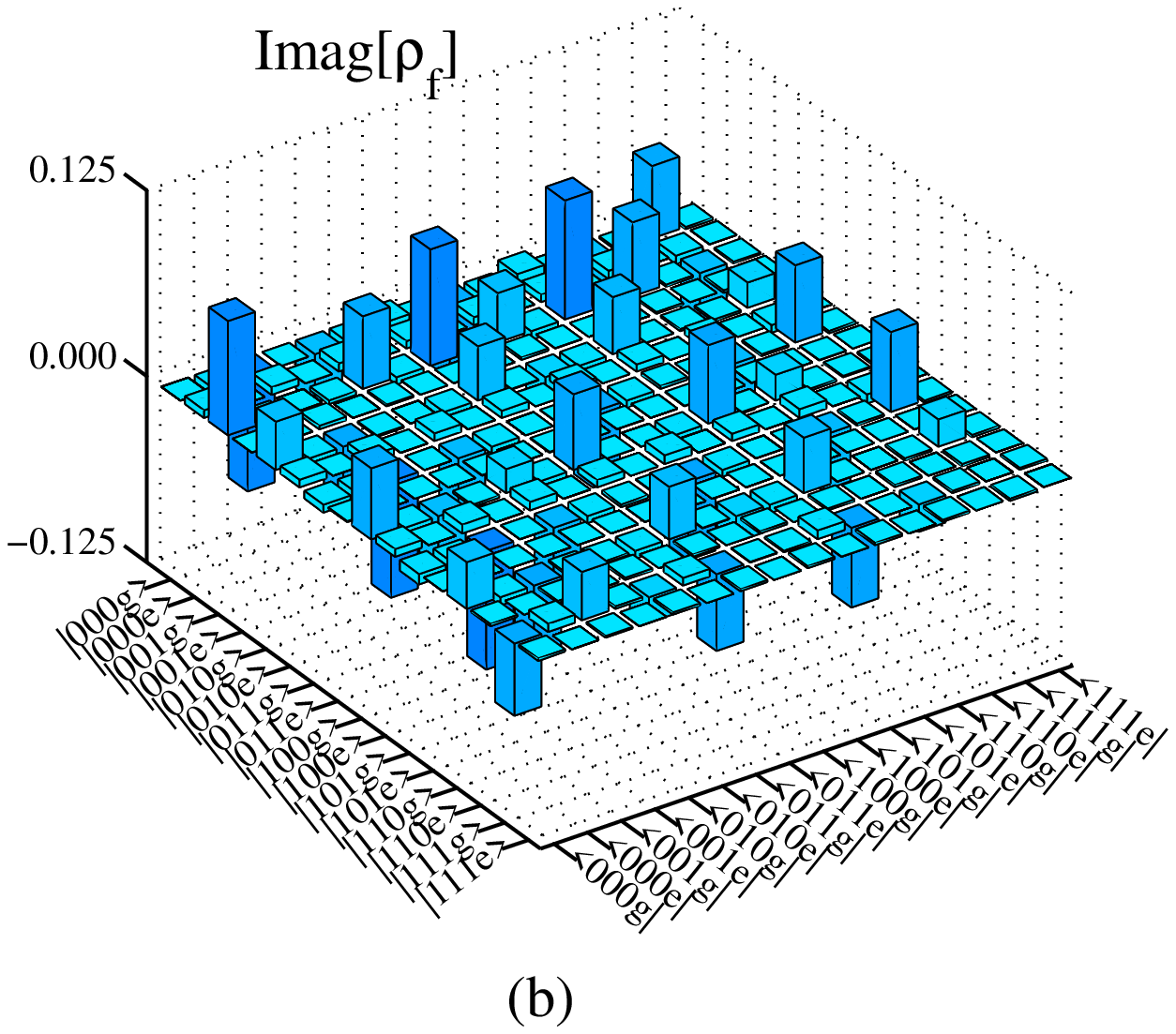}
\end{center}
\caption{(Color online)  The real part (a) and the imaginary part (b)
of the final state $\left\vert \Psi _{f}\right\rangle$ of the system
composed of the three microwave-photon resonators and the
superconducting qutrit in our cc-phase gate, respectively.}
\label{fig5}
\end{figure}

We simulate the evolution for the density operator of the system
with the initial state $\left\vert \Psi_{0}\right\rangle$, and the
reduced density operator of the final state $\left\vert \Psi_{f}\right\rangle $ shown in Eq.
(\ref{ccphasefinal}) is shown in Fig.\ref{fig5}. Here,
$\omega^{r_1}/(2\pi )=6.5GHz$, $\omega^{r_2}/(2\pi )=7.5GHz$,
$\omega^{r_3}/(2\pi )=7.5GHz$, $\omega^{g,e}/(2\pi
)=E_e-E_g=8.7GHz$, $\omega^{e,f}/(2\pi )=E_f-E_e=8.0GHz$,
$g^{g,e}_1/(2\pi )=g^{e,f}_1/(2\pi )=0.2GHz$, $g^{g,e}_2/(2\pi
)=g^{e,f}_2/(2\pi )=0.2GHz$, $g^{g,e}_3/(2\pi )=g^{e,f}_3/(2\pi
)=0.12GHz$, $\omega ^{d}/(2\pi )=8.1768GHz$, and $\Omega
=0.0266GHz$. From Fig.\ref{fig5}, one can see that the fidelity of
our cc-phase gate can reach $92.92\%$ within about 124.64 ns, without
considering the decoherence and leakage of the resonators.

\section{Discussion and summary}

The deterministic approaches to realize the nonlinear interaction
between two photons  for quantum computation are usually based on
the Kerr effect. Here we constructed  the local c-phase and cc-phase
gates on the resonator qudits in a microwave-photon quantum
processor assisted by only one transmon qutrit, resorting to the
combination of the number-state-dependent interactions between the
transmon qutrit and the resonator qudits and the simple resonant
interaction between the qutrit and one resonator qudit. Usually, the
processor on microwave-photon systems needs a tunable coupling
superconducting qubit or some tunable resonators
\cite{frederick,FWStrauch,Wu,Strauch}. The experiments showed that a
tunable coupling strength between a superconducting qubit and a
superconducting resonator is feasible
\cite{Laloy,Harris,Srinivasan}. Some recent experiments were
demonstrated for tuning the frequency of a resonator
\cite{Sandberg,Johansson,Ong}. In order to avoid shortening the
relaxation time of the qutrit, the processor needs some high-$Q$
resonators. That is, the present c-phase and cc-phase gates are
feasible, similar to those in Refs. \cite{frederick,Wu,Siyuan,Han}

In our calculation, the parameters of the transmon qutrit are chosen
as the same as those in Ref. \cite{Hoi}.  Actually, the coupling
strength for the two different transitions of a transmon qutrit and
a microwave-photon  resonator is asymptotically increased as $(E_{J}
/E_{C})^{1/4}$ (for a transmon qubit, $20<E_{J}
/E_{C}<5\times10^{4}$) \cite{Koch}. That is, it is reasonable to use
the same coupling strength for the two different transitions of a
transmon qutrit and a resonator for convenience.  The amplitudes of
the drive fields for constructing the c-phase and cc-phase gates are
too small (compared with the anharmonicity  between the two
transitions of the qutrit) to induce the influences coming from the
higher excited energy levels of the transmon qutrit.

The coherence time of a transmon qubit approaches  0.1 ms
\cite{Rigetti} and the life time of microwave photons contained in
resonators are always longer than that of a qutrit \cite{Devoret},
which means our gates can be operated several  hundreds of times
within the life time of the processor.  In order to evolve the
systems  from the dress states to the computational states in our
schemes for constructing the gates, one needs to tune on or off the
interaction between the resonators and the qutrit,  the same as in
Refs. \cite{frederick,Strauch}.  The quantum error coming from this
method in experiment is determined by the technique of the tunable
transition frequency of a superconducting resonator or the tunable
coupling strength between the qutrit and the resonators. In our
calculation, we don't consider the error coming from the preparation
of the initial states shown in Eqs.(\ref{cphase0}) and
(\ref{ccphase0}). To prepare the states shown in Eqs.(\ref{cphase0})
and (\ref{ccphase0}) from the state of the system composed of
multiple resonators coupled to a qutrit
$\otimes_i|0\rangle_i|g\rangle_q$, one needs to take the
single-qubit gate by appling a $\frac{\pi}{2}$ pulse on the qutrit,
and the state of the system is changed into
$\otimes_i\frac{1}{\sqrt{2}}|0\rangle_i(|g\rangle_q-i|e\rangle_q)$.
By resonating the qutrit and the resonator $j$ with the time
$t=\frac{3\pi}{2g_j^{g,e}}$, one can obtain  the state
$\otimes_{i,i\neq
j}\frac{1}{\sqrt{2}}(|0\rangle_j+|1\rangle_j)|0\rangle_i|g\rangle_q$
\cite{Wu}. By repeating the steps to the rest resonators, one can
obtain the states shown in Eqs.(\ref{cphase0}) and (\ref{ccphase0})
$\otimes_i\frac{1}{\sqrt{2^i}}(|0\rangle_i+|1\rangle_i)|g\rangle_q$.
The single-qubit gate can be realized with the quantum error of
$0.007 \pm 0.005$ \cite{singlegate} and even smaller than 0.0009
\cite{error}, which is too small to influence the fidelity of our
gates. By using a qubit to read out the state of the resonator
\cite{Hofheinz,Hofheinz2,Johnson}, one can complete fully the
tomography of the resonator logic gates \cite{FWStrauch}.

Now, let us compare our c-phase gate on two resonators with the one
constructed in Ref. \cite{FWStrauch}. In Ref. \cite{FWStrauch},
Strauch presented an interesting scheme to construct the c-phase
gate on two superconducting resonator qudits. In his work, each of
two resonators (A and B) is coupled to an auxiliary three-level
transmon or phase qutrit (a and b), and each qutrit should be
coupled to each other directly. Moreover, the c-phase gate on two
resonators based on the Fock states $|0\rangle$ and $|1\rangle$ is
constructed by first using the number-state-dependent interactions
between a superconducting qutrit and a resonator qudit (A to a and B
to b) twice, and then turning on the interaction between two
superconducting qutrits. Finally, the gate is completed by repeating
the first step. The operation time of this gate is 150 ns. In our
work, the c-phase gate is accomplished with two resonators which are
coupled to just one transmon qutrit. It is easier to extend our
two-resonator c-phase gate to a three-resonator processor assisted
by a transmon qutrit. The effects used for constructing our c-phase
gate on two resonators include the number-state-dependent
interaction between the qubit and one resonator-qudit subsystem, and
the simple resonant operation between the qubit and another
resonator-qudit subsystem. These different characteristics make us
get a higher fidelity c-phase gate with a faster operation time. The
fidelity of our gate is 99.51\% within the operation time of 93 ns.

Different from the effective c-phase gate constructed by Wu \emph{et
al.}  \cite{Wu} in which  both two resonators are  coupled to a
two-energy-level charge qubit by the number-state-dependent
interactions and it is completed without considering the existence
of the third energy level of the charge qubit (the operation time of
this gate is 125 ns), our c-phase gate on the two resonators (1 and
2) is accomplished by combination of the photon-number-dependent
frequency-shift effect on the transmon qutrit by the first resonator
and the simple resonant operation between the qutrit and the second
resonator. That is, resonator 2 is used to resonate with the qutrit
and resonator 1 is used to complete the selective rotation on the
qutrit by using the effect that the transition frequency of the
qutrit is determinated by the photon number in only  resonator 1,
which is simpler than the effect used in Ref. \cite{Wu}. This
different physical mechanism makes us obtain the higher fidelity and
faster c-phase gate on the two resonators.  Moreover, there are no
works about the construction of the cc-phase gate on three
microwave-photon-resonator qudits. The different devices in our work
make it possible to construct the cc-phase gate on three resonators,
far different from the previous proposals \cite{FWStrauch,Wu}. The
fidelity of our cc-phase gate is $92.92\%$ within the operation time
of 124.64 ns.

In summary, we have constructed  two  universal quantum gates, i.e.,
the c-phase and cc-phase gates in a microwave-photon quantum
processor which contains multiple superconducting
microwave-photon-resonator qudits coupled to  a $\Xi$-type
transmon qutrit. Our gates are based on the combination of
the number-state-dependent interaction between a transmon
qutrit and a resonator-qudit subsystem and the simple resonant operation
between the qutrit and another resonator-qudit subsystem, and they
have a high fidelity in a short operation time.  The algorithms of
our gates are based on the Fock states of the resonators, and the
microwave photon number in each resonator is limited to none or just
one.  Our universal quantum processor can deal with the quantum computation with microwave photons in resonators.

It is worth pointing out that the techniques for catching and
releasing  microwave-photon states from  a resonator to the
transmission line \cite{Yin} and the single-photon router in the
microwave regime \cite{Hoi} have been realized in experiments. The
microwave-photon quantum processor can act as an important platform
for quantum communication as well.

\section*{Acknowledgments}

This work is supported by the National Natural Science Foundation of
China under Grant No. 11174039 and  NECT-11-0031.


\end{document}